\documentclass[preprintnumbers,nofootinbib,noshowpacs,eqsecnum,superscriptaddress,prd]{revtex4}
\pdfoutput=1
\usepackage{graphicx}
\usepackage{amsmath,amsthm,amssymb}
\usepackage{mathrsfs,bbm}
\usepackage{array,color}
\usepackage{slashed}
\usepackage{tabularx}

\makeatletter       

\renewcommand{\p@subsection}{}

\makeatother

\newcommand*{\tvec}[1]{\boldsymbol{#1}}              
\newcommand*{\trans}{\mathrm{T}}                     

\DeclareMathOperator{\sgn}{sgn}			

\begin{document}
\title{Application of the Feynman-tree theorem together with BCFW recursion relations}

\author{M. Maniatis}
    \email[E-mail: ]{maniatis8@gmail.com}
    
\affiliation{Departamento de Ciencias B\'a{}sicas, 
UBB, Casilla 447, Chill\'a{}n, Chile.}

\begin{abstract}
Recently, it has been shown that on-shell scattering amplitudes
can be constructed by the Feynman-tree theorem combined with
the BCFW recursion relations. Since the BCFW relations are restricted
to tree diagrams, the preceding application of the Feynman-tree theorem
is essential. In this way amplitudes can be constructed
by on-shell and gauge-invariant tree amplitudes. Here we want to apply
this method to the electron-photon vertex correction.
We present all the single, double, and triple phase-space tensor integrals
explicitly and show that the sum of amplitudes coincides with the result of the
conventional calculation of a virtual loop correction. 
\end{abstract}

\maketitle

\section{Introduction}

The Feynman-tree theorem has been introduced by R. P. Feynman decades ago 
\cite{Feynman:1963ax, Feynman:FTT}. 
The  idea is to consider besides the
usual Feynman propagators $G_F(p)$ {\em advanced} propagators $G_A(p)$,
\begin{equation} \label{advanced}
G_F(p)  = \frac{i}{p^2 - m^2 + i \epsilon}, \qquad
G_A(p)  = \frac{i}{p^2 - m^2 - i \epsilon \sgn(p_0)}.
\end{equation}
With help of the identity
$\frac{1}{x\pm i \epsilon} = P.V. \bigg(\frac{1}{x}\bigg) \mp i \pi \delta(x)$,
where $P.V.$ is the principal value prescription we have
\begin{equation} \label{feynmanadvanced}
G_A(p) = G_F(p) - 2 \pi \;\delta^{(+)}(p^2 - m^2)
\end{equation}
with $\delta^{(+)}(p^2 - m^2) = \theta(p_0)\delta(p^2 - m^2)$, as usual.
Consider a loop 
with the
usual Feynman propagators $G_F(p)$ replaced by
the advanced propagators $G_A(p)$. In the loop momentum integration 
the poles of the
zero component lie therefore all above the real axis. 
Closing  the integration contour
in the lower half plane, we get the Feynman-tree theorem:
\begin{equation} \label{FeynmanTT}
\begin{split}
0 =& \int \frac{d^4 q}{(2\pi)^4}  N(q)\; \prod_i G_A^{(i)}(q-p_1-\ldots-p_i)\\ 
=& 
\int \frac{d^4 q}{(2\pi)^4}  N(q) \; \prod_i \bigg\{ G_F^{(i)}(q-p_1-\ldots-p_i) -
2 \pi \; \delta^{(+)}((q-p_1-\ldots-p_i)^2 -m^2) \bigg\}.
\end{split}
\end{equation}
Expanding the product on the right-hand side of the last equation,
the loop diagram is expressed in terms
of cut diagrams, where each cut is given by the corresponding delta distribution.
The function $N(q)$ denotes the numerator of the loop argument which in
general depends on the loop momentum. 
Recursively, all loops can be opened in this way. In each recursion
step a loop of $n$ propagators gives $2^n-1$ cut diagrams. In practical calculations
many of the cut diagrams vanish. Note that in each recursion step the loop 
order is decreased by at least one unit. Recursive application therefore represents any
loop diagram in terms of tree diagrams. 
In recent years there has been some interest in the Feynman-tree theorem; see for instance
\cite{Brandhuber:2005kd, Catani:2008xa, CaronHuot:2010zt, Bierenbaum:2010cy}.
Here we will not consider any modification of the initial theorem. In particular we 
do not want to restrict the statements to any specific loop order and merely keep the theorem
in its general form. 

In context with the cuts of the Feynman-tree theorem let us also mention
the generalized unitarity method; for a review see for instance \cite{Bern:2011qt}.
The idea of the unitarity method is to construct the amplitude as
 a function which coincides at all possible cuts.
The cut diagrams correspond to known diagrams or are at least diagrams which are easier to calculate.
In practical calculations the generalized unitarity method turns out to be 
very powerful. In contrast, the Feynman-tree theorem expresses any diagram 
systematically in terms of tree diagrams.

Recently it has been argued that the application of the Feynman-tree theorem
followed by BCFW recursion relations \cite{Britto:2004ap,Britto:2005fq} 
gives a systematic way to compute on-shell scattering
amplitudes \cite{Maniatis:2015kex,Maniatis:2016gui}. The BCFW recursion
relations split tree diagrams with external on-shell particles into on-shell amplitudes, 
{\em Opening the loops} recursively with the help
of the Feynman-tree theorem followed by BCFW recrusion relations,
on-shell and gauge-invariant scattering amplitudes arise in a natural way. 
Note that neither the Feynman-tree theorem nor
the BCFW recursion relation are restricted to four dimensions. Therefore, dimensional
regularization can be applied to handle the infrared and ultraviolet singularities. 
We emphasize that both methods are also not limited to massless particles. Calculations
in four dimensions with massless particles have only the advantage, that the
Weyl spinor formalism is directly applicable.

In general, in the analytic continuation of the external momenta
of the BCFW recursion relations, there may appear non-vanishing boundary terms. 
However, it has been shown that
in gauge theories and gravity this is not the case 
\cite{ArkaniHamed:2008yf,Feng:2009ei}. 

\begin{figure}[t]
\includegraphics[width=0.3\textwidth]{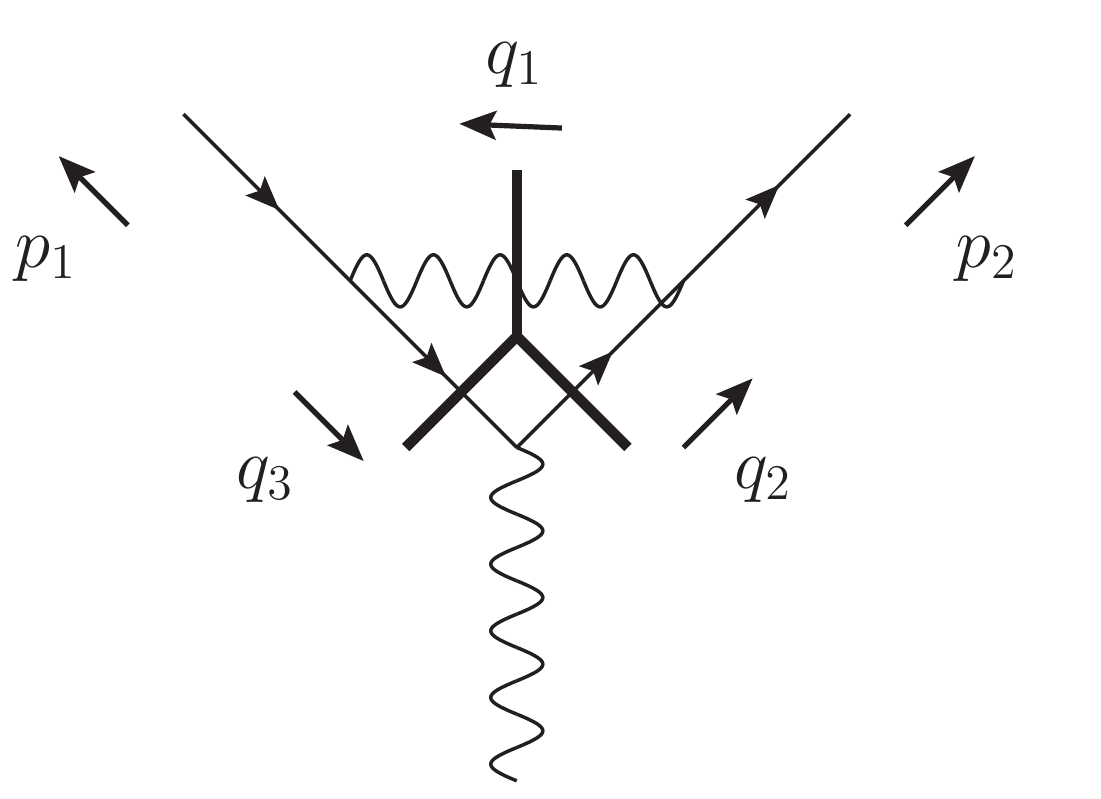}
\caption{\label{vertex}The cuts of the three propagators of
the electron-photon vertex correction. In the Feynman-tree theorem
all 7 possible cut combinations have to be applied. The
cut propagators correspond to pairs of unobserved particles.  Following 
BCFW these diagrams are given in terms of on-shell amplitudes.
}
\end{figure}

We consider here the explicit calculation of the 
vertex correction at next-to-leading order in the electromagnetic coupling $e$
for the electron-photon interaction as depicted in Fig. \ref{vertex}.
Since this vertex correction involves three propagators, the application of 
the Feynman-tree theorem gives $2^3-1=7$ diagrams; three single-cut 
diagrams, three double-cut diagrams, and one triple-cut diagram. These
diagrams can be computed by on-shell amplitudes following BCFW; see \cite{Maniatis:2016gui}
for details. 
In this paper we shall explicitly calculate all these amplitudes. We have to 
integrate over the phase space of the corresponding pairs of unobserved particles. 
The single-cut diagrams have been calculated in the case of
scalars \cite{Hernandez-Pinto:2015ysa}. Here
we consider all cut diagrams and extend these calculations to tensor integrals, which appear
in the calculation.
We will demonstrate that the phase-space integrations of cut diagrams can be
performed and that the sum of all amplitudes yields the well-known result. 
We emphasize that the essential building blocks are 
gauge-invariant on-shell amplitudes.


\section{Calculation}
\label{calculation}

Here we present the detailed calculation of the
different cut amplitudes contributing to the electron-photon vertex 
at next-to-leading order in the electromagnetic coupling $e$.

For simplicity, we perform the calculation without considering the mass of the electron.
This simplifies the calculations but does not mean any limitation of the method.
The phase-space integrations over the unobserved on-shell particle momenta are in general divergent, 
due to infrared and ultraviolet singularities.
Therefore we will regularize the integrations dimensional in $D= 4- 2 \epsilon$
dimensions, regularizing both, infrared and ultraviolet singularities. 

We begin with some general remarks on the phase-space integrations we are going to perform.
Let us consider an explicit cut integral, for instance the amplitude coming from the single cut
of the photon propagator with momentum $q_1$ (see Fig. \ref{vertex}), or equivalently, the amplitude
arising from the integration 
over an unobserved on-shell photon,
\begin{equation} \label{A1}
A_1(p_1, s_1, p_2, s_2,\lambda) = -e^3 \mu^{4-D} \int \frac{d^D l}{(2 \pi)^{D}}\;
2 \pi \; \delta^{(+)}(q_1^2)\; 
\frac{\bar{u}(p_2,s_2) \gamma_\alpha \slashed{q}_2\; \slashed{\epsilon}(\lambda)\; \slashed{q}_3 \gamma^\alpha u(p_1,s_1)}
{(q_2^2 + i\epsilon) (q_3^2+i\epsilon)} .
\end{equation}
The minus sign together with the delta distribution originate from the Feynman-tree theorem.  
The momentum and the spin of the external electrons are denoted by $p_{1/2}$ and $s_{1/2}$, respectively, whereas
$\lambda$ denotes the polarization of the external photon. 
The (cut) propagator momenta in terms of the integration momentum $l$ are
$q_1= l$, $q_2= l+p_2$, $q_3= l - p_1$.
Since the numerator is not modified by the Feynman-tree theorem (apart from including delta distributions accompanied
by $2\pi$ factors) we encounter in all 
amplitudes the same numerator
$
\gamma_\alpha \slashed{q}_2 \gamma_\mu
\slashed{q}_3 \gamma^\alpha =
q_2^\alpha q_3^\beta 
(
-2\gamma_\beta \gamma_\mu \gamma_\alpha
+(4-D) \gamma_\alpha \gamma_\mu \gamma_\beta
)
$, 
giving tensor integrals of first and second order. 

We use the parametrization of the on-shell loop momentum $q_i$ as suggested in
\cite{Hernandez-Pinto:2015ysa}. In that work also all
 scalar single cut integrals were presented. The explicit parametrization of the
on-shell loop momentum reads \cite{Hernandez-Pinto:2015ysa}
\begin{equation} \label{looppara}
q_i = \frac{\sqrt{s_{12}}}{2} \xi_i 
\left( 1, 2 \sqrt{v_i (1-v_i)} \tvec{e}_{i, \trans}, 1- 2 v_i \right)^\trans
\end{equation}
with $\xi_i \in [0, \infty [$, $v_i \in [0,1]$ and $\tvec{e}_{i, \trans}$ a unit
vector in transverse direction.
In this parametrization we have $q_i^2=0$ and with $\tvec{p}_1$ pointing in positive $z$ direction
as well as $\tvec{p}_2 = - \tvec{p}_1$,
\begin{equation}
2 q_i p_1 = s_{12}\; \xi_i\; v_i, \quad
2 q_i p_2 = s_{12}\; \xi_i\; (1 - v_i)\quad {\text{with}}\quad s_{12}=(p_1+p_2)^2.
\end{equation}
Shifting the phase-space momentum $l$ to one of the cut momenta we have
\begin{equation}
d^D l = d^D q_i
= d q_{i, 0}\;  d^{D-1} \tvec{q}_i = d q_{i, 0}\;  d|\tvec{q}_i|\; |\tvec{q}_i|^{D-2}\; d \Omega_{D-1}
\end{equation}
with $\Omega_D$ the solid angle in $D$ dimensions.
With $|\tvec{q}_i| = q_{i, 0} = \frac{\sqrt{s_{12}}}{2} \xi_i$ as well as 
$v_i = 1/2 (1- \cos (\theta_1))$ we get 
\begin{equation} \label{solid}
d \Omega_{D-1}  = \Omega_{D-2} 2^{D-3}\; (v_i(1-v_i))^{(D-4)/2}\; d v_i,
\quad
\Omega_{D-2} = \frac{2 \cdot \pi^{(D-2)/2} } {\Gamma((D-2)/2)}.
\end{equation}
We have seen that we have to deal with tensor integrals, arising from the numerator of the 
integrand in \eqref{A1}. Analogously to the tensor reduction of 
standard integrals \cite{Passarino:1978jh} we do a covariant decomposition with respect to
the available momenta $p_1$ and $p_2$, explicitly,
\begin{equation} \label{TRdec}
I_j^\alpha  = p_1^\alpha C_1 + p_2^\alpha C_2, \qquad
I_j^{\alpha \beta} =
g^{\alpha \beta} C_{00}
+
\sum_{i,k=1}^2
p_i^\alpha p_k^\beta C_{ik} .
\end{equation}
We contract these covariant decompositions 
with $p_{1, \alpha}$ and $p_{2, \alpha}$ in the
case of the tensors of order one, and with $g_{\alpha \beta}$, 
$p_{1, \alpha} p_{1, \beta}$, $p_{2, \alpha} p_{2, \beta}$, $p_{1, \alpha} p_{2, \beta}$, 
$p_{2, \alpha} p_{1, \beta}$, respectively,
in case of the tensors of second order. Computing these invariants explicitly we 
determine all the coefficients in \eqref{TRdec} and in this way the tensor integrals.


With these preparations let us calculate the amplitude $A_1$, \eqref{A1}, corresponding to a single photon-propagator cut.
Let us begin with the calculation of the scalar integral,
\begin{equation}
I_1 = - \mu^{4-D} \int \frac{d^D l}{(2 \pi)^{D}}\;
2 \pi \delta^{(+)}(q_1^2)\; \frac{1}{(q_2^2+i\epsilon)(q_3^2+i\epsilon)}.
\end{equation}
We shift the integration variable to the cut momentum $q_1$.
Integration over the zero component of $q_1$ justifies the on-shell
parametrization \eqref{looppara}.
With $q_2= q_1 + p_2$ and $q_3= q_1-p_1$,
we have $q_2^2 = -2 (q_1 p_2) = - s_{12} \xi_1 (1 - v_1)$,
$q_3^2 = -2 (q_1 p_1) = - s_{12} \xi_1 v_1$, and get \cite{Hernandez-Pinto:2015ysa}
\begin{equation} 
I_1 =
\Omega_{D-2}\;
\frac{\mu^{4-D}}{4(2 \pi)^{D-1}} s_{12}^{D/2-3}
\int
d \xi_1 \; 
\xi_1^{D-5}   d v_1 \frac{(v_1 (1 - v_1))^{(D-4)/2}}{v_1 (1 - v_1)} = 0.
\end{equation}
The  integration  over $\xi_1$ in $D$ dimensions is {\em scaleless} and therefore has to vanish; 
for details see for instance \cite{Pak:2010pt}.
Computing  the invariants from \eqref{TRdec} we see that we always get {\em scaleless}
integrals and therefore we find $I_1^\alpha =0$ and  $I_1^{\alpha \beta} = 0$. The photon single-cut amplitude
gives no contribution here,
\begin{equation}
A_1(p_1, s_1, p_2, s_2,\lambda) = 0 .
\end{equation}


We proceed with the single-cut amplitude integrating over the phase space of the
on-shell electron with momentum $q_2$, 
\begin{equation} \label{A2}
A_2(p_1, s_1, p_2, s_2,\lambda) = -e^3 \mu^{4-D} \int \frac{d^D l}{(2 \pi)^{D}}\;
2 \pi \; \delta^{(+)}(q_2^2)\; 
\frac{\bar{u}(p_2,s_2) \gamma_\alpha \slashed{q}_2 \slashed{\epsilon}(\lambda) \slashed{q}_3 \gamma^\alpha u(p_1,s_1)}
{(q_1^2 + i\epsilon) (q_3^2+i\epsilon)} .
\end{equation}
We shift the integration to $q_2$ and achieve with 
$q_1^2 = -s_{12} \xi_2(1-v_2)$ and $q_3^2= s_{12}(1-\xi_2)$
for the scalar integral
\begin{equation} \begin{split}
I_2 = &  -\mu^{4-D} \int \frac{d^D l}{(2 \pi)^{D}}\;
2 \pi \; \delta^{(+)}(q_2^2)\; \frac{1}{(q_1^2+i\epsilon) (q_3^2+i\epsilon)}
\\ =& 
\frac{\mu^{4-D}}{4(2 \pi)^{D-1}} s_{12}^{D/2-3}
\Omega_{D-2}
\int
d \xi_2 
\frac{\xi_2^{D-4}}{1-\xi_2+ i \epsilon}  
\; d v_2\; 
\frac{(v_2(1-v_2))^{(D-4)/2}}{1 - v_2} .
\end{split}
\end{equation}
The integrals over $\xi_i$ and $v_i$ can be found in the appendix. 
We proceed with the tensor integrals \eqref{TRdec}. Decomposing in the
numerator of \eqref{A2}
$q_2^\alpha q_3^\beta = q_2^\alpha q_2^\beta - q_2^\alpha (p_1 + p_2)^\beta$
we compute the invariants
\begin{equation} \begin{split}
p_{1, \alpha}  I_2^\alpha = &
\frac{1}{2} 
\frac{\mu^{4-D}}{4(2 \pi)^{D-1}}
 s_{12}^{D/2-2} \Omega_{D-2}
 \int d \xi_2 d v_2
\frac{\xi_2^{D-3}}{(1-\xi_2)} \frac{v_2}{(1-v_2)}(v_2 (1-v_2))^{(D-4)/2},
\\
p_{2, \alpha}  I_2^\alpha = &
\frac{1}{2} 
\frac{\mu^{4-D}}{4(2 \pi)^{D-1}}
 s_{12}^{D/2-2} \Omega_{D-2}
 \int d \xi_2 d v_2
\frac{\xi_2^{D-3}}{(1-\xi_2)} (v_2 (1-v_2))^{(D-4)/2} .
\end{split} \end{equation}
With the help of the integrals in the appendix we get the coefficients $C_1$ and $C_2$ 
in \eqref{TRdec}, that is, $I_2^\alpha$.

For the second-order tensor we compute all the corresponding invariants,
\begin{equation} \begin{split}
&
g_{\alpha  \beta} I_2^{\alpha \beta} = 0,\\
&
p_{1, \alpha} p_{1, \beta} I_2^{\alpha \beta} = 
\frac{1}{4} \frac{\mu^{4-D}}{4(2 \pi)^{D-1}}
 s_{12}^{D/2-1} \Omega_{D-2}
 \int d \xi_2 d v_2
\frac{\xi_2^{D-2}}{(1-\xi_2)} \frac{v_2^2}{(1-v_2)}(v_2 (1-v_2))^{(D-4)/2},\\
&
p_{2, \alpha} p_{2, \beta} I_2^{\alpha \beta} = 
\frac{1}{4} \frac{\mu^{4-D}}{4(2 \pi)^{D-1}}
 s_{12}^{D/2-1} \Omega_{D-2} 
 \int d \xi_2 d v_2
\frac{\xi_2^{D-2}}{(1-\xi_2)} (1-v_2)(v_2 (1-v_2))^{(D-4)/2},\\
&
p_{1, \alpha} p_{2, \beta} I_2^{\alpha \beta} =
p_{2, \alpha} p_{1, \beta} I_2^{\alpha \beta}
= 
\frac{1}{4} \frac{\mu^{4-D}}{4(2 \pi)^{D-1}}
 s_{12}^{D/2-1} \Omega_{D-2} 
 \int d \xi_2 d v_2
\frac{\xi_2^{D-2}}{(1-\xi_2)} v_2(v_2 (1-v_2))^{(D-4)/2} .
\end{split}\end{equation}
With the explicit $\xi_i$ and $v_i$ integrations in the appendix
we construct the tensor integral $I_2^{\alpha \beta}$.
In this way we get the amplitude,
\begin{equation} \label{A2res}
\qquad A_2(p_1, s_1, p_2, s_2,\lambda) =
e^3
\bar{u}(p_1,s_1)\slashed{\epsilon}(\lambda)u(p_2,s_2) 
\frac{2 \pi \mu^{2\epsilon}}{4(2\pi)^{3-2\epsilon}}
s_{12}^{-\epsilon}\;
\Omega_{D-2}
\frac{\Gamma(2-\epsilon)\Gamma(-\epsilon)}{\Gamma(3-2\epsilon)}
\left(\frac{1}{\tan(2 \pi \epsilon)}-i\right)
(2-\epsilon +2 \epsilon^2) ,
\end{equation}
with the solid angle $\Omega_{D-2}$ given in \eqref{solid}.

The amplitude $A_3$ reads
\begin{equation} \label{A3}
A_3(p_1, s_1, p_2, s_2,\lambda) = -e^3 \mu^{4-D} \int \frac{d^D l}{(2 \pi)^{D}}\;
2 \pi \; \delta^{(+)}(q_3^2)\; 
\frac{\bar{u}(p_2,s_2) \gamma_\alpha \slashed{q}_2 \slashed{\epsilon} \slashed{q}_3 \gamma^\alpha u(p_1,s_1)}
{(q_1^2 + i\epsilon) (q_2^2+i\epsilon)}
\end{equation}
and we compute the scalar integral (see also \cite{Hernandez-Pinto:2015ysa})
\begin{equation} \begin{split}
I_3 =& -\mu^{4-D} \int \frac{d^D l}{(2 \pi)^{D}}\;
2 \pi\; \delta^{(+)}(q_3^2)\; \frac{1}{(q_1^2+i\epsilon) (q_2^2+i\epsilon)}
\\ = &
-
\frac{\mu^{4-D}}{4(2 \pi)^{D-1}} s_{12}^{D/2-3}
\Omega_{D-2}
\int
d \xi_3 \; d v_3\; 
\frac{\xi_3^{D-4}}{1+\xi_3} 
\frac{(v_3(1-v_3))^{(D-4)/2}}{v_3} ,
\end{split} \end{equation}
where the integration is shifted to $q_3$, with $q_1 = q_3 + p_1$ and
$q_2 = q_3 + p_1 + p_2$, that is,
$q_1^2 = 2 q_3 p_1=  s_{12} \xi_3 v_3$ and $q_2^2=s_{12}(1+\xi_3)$. The tensor
in the numerator \eqref{A3} becomes
$q_2^\alpha q_3^\beta 
=
q_3^\alpha q_3^\beta + (p_1+p_2)^\alpha q_3^\beta$.
We compute the invariants
\begin{equation} \begin{split}
p_{1, \alpha}  I_3^\alpha = &
-\frac{1}{2} 
\frac{\mu^{4-D}}{4(2 \pi)^{D-1}}
 s_{12}^{D/2-2} \Omega_{D-2}
 \int d \xi_3 d v_3
\frac{\xi_3^{D-3}}{(1+\xi_3)} (v_3 (1-v_3))^{(D-4)/2},
\\
p_{2, \alpha}  I_3^\alpha = &
-\frac{1}{2} 
\frac{\mu^{4-D}}{4(2 \pi)^{D-1}}
 s_{12}^{D/2-2} \Omega_{D-2}
 \int d \xi_3 d v_3
\frac{\xi_3^{D-3}}{(1+\xi_3)} \frac{1-v_3}{v_3} (v_3 (1-v_3))^{(D-4)/2} .
\end{split} \end{equation}
With the help of the integrals in the appendix we get the coefficients in \eqref{TRdec},
hence, $I_3^\alpha$.

The invariants for the second-order tensor read
\begin{equation}\begin{split}
&
g_{\alpha  \beta} I_3^{\alpha \beta} = 0,\\
&
p_{1, \alpha} p_{1, \beta} I_3^{\alpha \beta} =
-\frac{1}{4} \frac{\mu^{4-D}}{4(2 \pi)^{D-1}}
 s_{12}^{D/2-1} \Omega_{D-2} 
 \int d \xi_3 d v_3
\frac{\xi_3^{D-2}}{(1+\xi_3)} v_3\; (v_3 (1-v_3))^{(D-4)/2},\\
&
p_{2, \alpha} p_{2, \beta} I_3^{\alpha \beta} =
-\frac{1}{4} \frac{\mu^{4-D}}{4(2 \pi)^{D-1}}
 s_{12}^{D/2-1} \Omega_{D-2} 
 \int d \xi_3 d v_3
\frac{\xi_3^{D-2}}{(1+\xi_3)} \frac{(1-v_3)^2}{v_3}\; (v_3 (1-v_3))^{(D-4)/2},\\
&
p_{1, \alpha} p_{2, \beta} I_3^{\alpha \beta} = 
-\frac{1}{4} \frac{\mu^{4-D}}{4(2 \pi)^{D-1}}
 s_{12}^{D/2-1} \Omega_{D-2} 
 \int d \xi_3 d v_3
\frac{\xi_3^{D-2}}{(1+\xi_3)} (1-v_3)\; (v_3 (1-v_3))^{(D-4)/2}.
\end{split}\end{equation}
We construct the tensor integral $I_3^{\alpha \beta}$ from these invariants
and get the amplitude,
\begin{equation} \label{A3res}
\qquad A_3(p_1, s_1, p_2, s_2,\lambda) =
e^3
\bar{u}(p_1,s_1)\slashed{\epsilon}(\lambda)u(p_2,s_2) \cdot
\frac{2^{2\epsilon}\pi^{3/2} \mu^{2\epsilon}}{8(2\pi)^{3-2\epsilon}}
\;
s_{12}^{-\epsilon}
\;
\Omega_{D-2} 
\frac{\Gamma(-\epsilon)}{\Gamma(3/2-\epsilon)}
\frac{1}{\sin(2 \pi \epsilon)}
(2-\epsilon +2 \epsilon^2) .
\end{equation}

We proceed with the double-cut amplitude
\begin{equation} \label{A23}
A_{23}(p_1, s_1, p_2, s_2,\lambda) = e^3 \mu^{4-D} \int \frac{d^D l}{(2 \pi)^{D}}\;
2 \pi \; \delta^{(+)}(q_2^2)\; 
2 \pi \; \delta^{(+)}(q_3^2)\; 
\frac{\bar{u}(p_2,s_2) \gamma_\alpha \slashed{q}_2 \slashed{\epsilon} \slashed{q}_3 \gamma^\alpha u(p_1,s_1)}
{q_1^2 + i\epsilon} .
\end{equation}
The corresponding scalar integral is, shifting the integration to $q_3$,
\begin{equation} \label{I23} \begin{split}
I_{23} =& \mu^{4-D} \int \frac{d^D l}{(2 \pi)^{D}}\;
2 \pi  \delta^{(+)}(q_2^2)\;2 \pi \delta^{(+)}(q_3^2) \frac{1}{q_1^2+i\epsilon}
\\ = &
\frac{\mu^{4-D}}{(2 \pi)^{D-2}} 
\frac{s_{12}^{D/2-3}}{2^{D-1}} 
\int
d \xi_3 \; d \Omega_{D-1}\; \; \delta^{(+)}(1+\xi_3)
\frac{\xi_3^{D-4}}{v_3} 
\\ = &
\;0 .
\end{split} \end{equation}
The last step follows from the integration over $\xi_i \in [0, \infty]$ and the delta distribution.
This result holds also for the corresponding tensor integrals and we find
\begin{equation}
A_{23}(p_1, s_1, p_2, s_2,\lambda) = 0.
\end{equation}

All the other double-cut amplitudes and the triple-cut amplitude vanish, too. 
In particular, the two remaining double-cut amplitudes turn out to be scaleless
and for the triple-cut amplitude, the corresponding scalar integral
\begin{equation}
I_{123} = \mu^{4-D} \int \frac{d^D l}{(2 \pi)^{D}}\;
2 \pi\; \delta^{(+)}(q_1^2)\;
2 \pi\; \delta^{(+)}(q_2^2)\;
2 \pi\; \delta^{(+)}(q_3^2)
=
0
\end{equation}
vanishes since the delta distributions can not be simultaneously fulfilled, 
like in \eqref{I23}. This holds also for the corresponding tensor integrals. 

The complete amplitude for the electron-photon vertex correction is
therefore the sum of \eqref{A2res} and \eqref{A3res}, expanding around $\epsilon=0$,
\begin{multline}
A(p_1, s_1, p_2, s_2,\lambda) =
\frac{e^3}{4}
\bar{u}(p_1,s_1)\slashed{\epsilon}(\lambda)u(p_2,s_2) 
\frac{\Omega_{D-2} }{(2\pi)^{3-2\epsilon}}
\cdot\\
\left\{
-\frac{2}{\epsilon^2}
+\frac{1}{\epsilon} \left( 2 \ln \left( \frac{-s_{12}}{\mu^2} \right) - 3 \right)
	- \ln^2 \left( \frac{-s_{12}}{\mu^2} \right)  + 3 \ln \left( \frac{-s_{12}}{\mu^2} \right)  - 8
\right\}
+ {\cal O}(\epsilon) ,
\end{multline}
which is the well-known result we get in conventional computation from a virtual
Feynman diagram loop.



\section{Conclusions}
The Feynman-tree theorem on the one hand opens recursively all loops of a 
diagram yielding tree
diagrams and on the other hand
the BCFW recursion relations transform the tree diagrams into on-shell
amplitudes. In this way Feynman diagrams can be decomposed into gauge-invariant
on-shell amplitudes. We have considered an explicit example, the electron-photon
vertex correction in QED. We encounter phase-space integrations over momenta of
unobserved particles arising form single, double, and triple cuts. 
The tensor integrals have been calculated explicitly. 
Even though the number of amplitudes originating from the Feynman-tree theorem 
is in general rather large, we have seen that often multi-cut amplitudes 
vanish due to the phase-space kinematics. 
We recovered the well-known result for the electron-photon vertex correction
from the calculation of the amplitudes. This demonstrates the consistency of
the method to combine the Feynman-tree theorem with BCFW recursion relations,
in a physical amplitude. 
We leave it open for future
work to do the calculation with massive particles. Moreover, it would
be interesting to show the explicit calculation at higher
perturbative order, corresponding
in terms of usual Feynman diagrams to 
higher loop order.

We expect that the combination of the Feynman-tree theorem with the BCFW
recursion relations gives some new insights into scattering amplitudes.
In particular, it is quite striking that Feynman diagrams in a gauge theory
decay into on-shell and gauge-invariant building blocks, which can 
be calculated separately.

\section*{Acknowledgement}
We would like to thank 
Simon Caron-Huot and Otto Nachtmann
for many valuable comments and suggestions.
This work is supported partly by the 
Chilean research project FONDECYT, with project number
1140568, as well as by the the UBB project GI-152609/VC {\em F\'{i}sica
de Altas Energias}.

\appendix

\section{Integrals}

Here we give all the elementary integrals needed in the calculations:
\begin{equation} \begin{split} 
&\int_0^\infty
d \xi
\frac{\xi^{2-2\epsilon}}{1-\xi+ i \epsilon}  
=
\int_0^\infty
d \xi
\frac{\xi^{1-2\epsilon}}{1-\xi+ i \epsilon}  
=
\int_0^\infty
d \xi
\frac{\xi^{-2\epsilon}}{1-\xi+ i \epsilon}  
=
-\int_{-\infty}^0
d \xi 
\frac{\xi^{-2\epsilon}}{1-\xi}  
=
-(-1)^{-2\epsilon}\frac{\pi}{\sin(2 \pi \epsilon)},\\
&\int_0^\infty
d \xi
\frac{\xi^{2-2\epsilon}}{1+\xi}  
=
-
\int_0^\infty
d \xi
\frac{\xi^{1-2\epsilon}}{1+\xi}  
=
\int_0^\infty
d \xi
\frac{\xi^{-2\epsilon}}{1+\xi}  
=
\frac{\pi}{\sin(2 \pi \epsilon)},\\
&\int_0^1 d v\; 
(v(1-v))^{-\epsilon}
=
\frac{\Gamma^2(1-\epsilon)}{\Gamma(2-2\epsilon)},\\
&\int_0^1 d v\; 
(v(1-v))^{-\epsilon} \frac{v}{1-v}
=
\int_0^1 d v\; 
(v(1-v))^{-\epsilon} \frac{1-v}{v}
=
\frac{\Gamma(2-\epsilon)\Gamma(-\epsilon)}{\Gamma(2-2\epsilon)},\\
&\int_0^1 d v\; 
\frac{(v(1-v))^{-\epsilon}}{1 - v}
=
\int_0^1 d v\; 
\frac{(v(1-v))^{-\epsilon}}{v}
=
\frac{\Gamma(1-\epsilon) \Gamma(-\epsilon)}{\Gamma(1-2\epsilon)},\\
&\int_0^1 d v\; 
(v(1-v))^{-\epsilon} \frac{v^2}{1 - v}
=
\frac{\Gamma(3-\epsilon) \Gamma(-\epsilon)}{\Gamma(3-2\epsilon)},\\
&\int_0^1 d v\; 
(v(1-v))^{-\epsilon} \frac{(1-v)^2}{v}
=
\frac{\sqrt{\pi}}{4}
4^{\epsilon}
(2-\epsilon)
\frac{\Gamma(-\epsilon)}{\Gamma(3/2-\epsilon)},\\
&\int_0^1 d v\; 
(v(1-v))^{-\epsilon} v
=
\frac{\Gamma(1-\epsilon) \Gamma(2-\epsilon)}{\Gamma(3-2\epsilon)},\\
&\int_0^1 d v\; 
(v(1-v))^{-\epsilon} (1-v)
=\frac{\sqrt{\pi}}{4}4^{\epsilon}
\frac{\Gamma(1-\epsilon)}{\Gamma(3/2-\epsilon)}.
\end{split}\end{equation}



\end{document}